\documentclass[10pt,conference]{IEEEtran}

\usepackage{oldlfont}
\usepackage{psfrag}
\usepackage{latexsym}
\usepackage{graphics}
\usepackage{epsfig}
\usepackage{amsmath}
\usepackage{amssymb}
\usepackage{array}

\begin{document}

\title{Entropy coding with Variable Length \\ Re-writing Systems}

\author{\authorblockN{Herve Jegou}
\authorblockA{University of Rennes\\
Email: Herve.Jegou@irisa.fr}
\and
\authorblockN{Christine Guillemot}
\authorblockA{IRISA/INRIA \\
Email: Christine.Guillemot@irisa.fr}
}


\maketitle


\def\A{{\mathcal A}}
\def\B{{\mathcal B}}
\def\C{{\mathcal C}}
\def\X{{\mathcal X}}
\def\R{{r}}
\def\Ri{{\mathcal R}_i}
\def\S{{\mathbf S}}
\def\s{{\mathbf s}}
\def\E{{\mathbf E}}
\def\e{{\mathbf e}}
\def\P{{\mathbb P}}
\def\b{\overline{b}}
\def\l{\overline{l}}
\def\voidsequence{\varepsilon}

\newcounter{example}
\setcounter{example}{0}

\newcounter{definition}
\setcounter{definition}{0}

\newcounter{property}
\setcounter{property}{0}

\def\theexample{\arabic{example}}
\def\thedefinition{\arabic{definition}}
\def\theproperty{\arabic{property}}

\newenvironment{example}%
{\refstepcounter{example} 
\vspace{0.2cm}
\noindent {\it Example \theexample:} \bgroup \sl}%
{\egroup \vspace{0.2cm}}

\newenvironment{definition}%
{\refstepcounter{definition} 
\vspace{0.2cm}
\noindent {\bf Definition \thedefinition:} \bgroup}%
{\egroup}

\newenvironment{property}%
{\refstepcounter{property} 
\vspace{0.2cm}
\noindent {\it Property \theproperty:} \bgroup \sl}%
{\egroup}

\begin{abstract} 
This paper describes a new set of block source codes well suited 
for data compression.
These codes are defined by sets of productions rules of the form $a \l \rightarrow \b$, 
where $a \in \A$ represents a value from the source alphabet $\A$ 
and $\l, \b$ are -small- sequences of bits. 
These codes naturally encompass other Variable Length Codes (VLCs) such as Huffman codes. 
It is shown that these codes may have a similar or even a shorter mean 
description length 
than Huffman codes for the same encoding and decoding complexity.
A first code design method allowing to preserve the lexicographic order 
in the bit domain is described. 
The corresponding codes have the same mean description length (mdl) 
as Huffman codes from which they are constructed. 
Therefore, they outperform from a compression point of view
the Hu-Tucker codes designed to offer 
the lexicographic property in the bit domain. 
A second construction method allows to obtain codes such that the marginal bit probability 
converges to 0.5 as the sequence length increases and this is achieved 
even if the probability distribution function is not known by the encoder.
\end{abstract}

\section{Introduction} 
\label{sec:introduction}

Grammars are powerful tools which are widely used in Computer Sciences. 
Most of lossless compression algorithms can actually be formalized 
with grammars.
Codes explicitly based on grammars have been considered as a mean 
for data compression \cite{KiY00}. 
These codes losslessly encode a sequence in two steps. 
A first analysis step consists in finding the production rules. 
A second step applies these rules to the sequence to be encoded. 
These codes have mainly been compared with dictionary-based compression algorithms 
such as LZ77 \cite{LZ77} or \cite{LZ78}, which also implicitly use 
the grammar formalism. 
All these codes have in common the fact that 
the set of production rules depends on the data to be encoded, 
and not only on the source properties.  

In this paper, a new set of codes based on specific production rules is introduced. 
In contrast with LZ77-like algorithms or grammar codes, 
the set of production rules is fixed. 
In contrast with grammar codes introduced so far in the literature, the codes 
described here encompass Huffman codes \cite{Huf52} 
(but not Variable-to-Fixed Length codes such as Tunstall Codes \cite{Tun67}). 
The form of the production rules is presented in Section~\ref{sec:notations}. 
The sequence of bits generated by a given production rule 
may be re-written by a subsequent production rule. 
They lead to the same encoding and decoding complexity as Huffman codes. 
A possible drawback of these codes would be that they require backward encoding. 
However, since most applications deal with block encoding, 
the forward encoding property is not absolutely required.
In Section~\ref{sec:encodingdecoding}, the decoding and encoding procedures 
with automata will be described. 
The compression efficiency of these codes will be analyzed 
in Section~\ref{sec:compressionefficiency}. 
It is shown in an example that the proposed codes 
allow for better compression efficiency than Huffman codes. 


Two code construction methods are then described.
The first method constructs a set of production rules preserving 
the lexicographic order of the original source sequence in the bit domain. 
This property is obviously of interest for database applications, 
since it allows to process comparative queries directly in the bit domain, 
hence avoiding the prematurate decoding of 
the compressed dictionary for the query itself. 
Note that the lexicographic VLC of minimal mdl
is usually obtained with the Hu-Tucker algorithm \cite{HuT71}. 
This algorithm is optimal in the set of VLCs. 
For some sources, the Hu-Tucker codes may have the same compression efficiency 
as Huffman codes, but it is not the case in general. 
The method proposed in Section~\ref{sec:lexi} constructs lexicographic codes 
with the same compression performance as Huffman codes and that allow 
for symbol per symbol encoding and decoding procedures. 
Obtaining together the properties of lexicographic order preservation and high 
compression efficiency illustrates the interest of codes based on the proposed 
set of production rules. 

The second construction method described in Section~\ref{sec:mirror} allows to obtain codes, 
for stationary sources, such that the marginal bit probability is equal to 0.5.
The main advantage of these codes is that this probability is equal to 0.5 even if 
the actual source probabilities are not known at the encoder, or if 
the assumed {\it a priori} probabilities differ from the true probabilities. 
Since channel encoders widely assume that 0{\it s} and 1{\it s} have the 
same probability, this property is of interest when compressed bitstreams 
protected by such encoders are transmitted over noisy channels.

\section{Problem statement and Notations}
\label{sec:notations}

In the sequel random variables are denoted by upper cases 
and the corresponding realizations are denoted by lower cases.
Sets are denoted by calligraphic characters. 
The cardinality of a given set $\X$ is denoted $|\X|$. 
We define $\X^+= \bigcup_{i=1}^{\infty} \X^i$ and $\X^* = \{\voidsequence\} \cup \X^+$, 
where $\voidsequence$ denotes the void sequence. 
Hence $\X^*$ denotes the set of sequences composed of elements of $\X$.
Let $\S \in \A^+$ be a sequence of source symbols taking 
their values in a finite alphabet $\A = \{a_1,\hdots a_i, \hdots \}$. 
The length of such a sequence is denoted $L(\S)$. 
The alphabet $\A$ is assumed to be ordered according to a total order $\prec$. 
Without loss of generality, we assume that $a_{1} \prec a_{2} \hdots \prec a_i \hdots \prec a_{|\A|}$.
Let us define $\B=\{0,1\}$. 
In the sequel, the emitted bitstream is denoted $\E=E_1 \dots E_{L(\E)} \in \B^*$ 
and its realization is denoted $\e=e_1 \dots e_{L(\e)}$.

\begin{definition}
\label{def:vlrs}
A Variable Length Re-writing System (VLRS) is a set 
${\mathcal R}=\bigcup_{i \in \A} {\mathcal R}_{i}$, 
where $\Ri$ denotes the set of rules related to a given symbol $a_i$, defined as
\begin{equation}
\nonumber
\begin{array}{lccc}
\R_{1,1}: & a_1 \, \l_{1,1} & \rightarrow & \b_{1,1} = b_{1,1}^1 \hdots b_{1,1}^{L(\b_{1,1})}, \\
~ & ~ & \vdots \\
\R_{i,j}: & a_i \, \l_{i,j} & \rightarrow & \b_{i,j} = b_{i,j}^1 \hdots b_{i,j}^{L(\b_{i,j})}, \\
~ & ~ & \vdots \\
\R_{|\A|,|{\mathcal R}_{|\A|}|}: & a_{|\A|} \, \l_{|\A|,|{\mathcal R}_{|\A|}|} 
& \rightarrow & \b_{|\A|,|{\mathcal R}_{|\A|}|}
\end{array}
\end{equation}
where $\l_{i,j} \in \B^*, \b_{i,j} \in \B^+$. This set is such that
\begin{enumerate}
\item $\forall i,\ |\Ri| \geq 1$,
\label{cond1}
\\
\item The set $\bigcup_{i=1}^{|\A|} \bigcup_{j=1}^{|\Ri|} \{ \b_{i,j} \}$ 
forms a prefix code (i.e. no codeword is the prefix of another \cite{CoT91_5}).
\label{cond2}
\\
\item $\forall i,\ \bigcup_{j=1}^{|\Ri|} \{ \l_{i,j} \}$ 
is the set $\{\voidsequence\}$ or forms a full prefix code 
(i.e, such that the Kraft sum is equal to $1$).
\label{cond3}
\\
\item $\forall i\ \forall i' \neq i, \forall j,j',\ \b_{i,j} 
= \l_{i',j'}$ 
or $\b_{i,j}$ is {\it not} a prefix of $\l_{i',j'}$.
\label{cond4}
\end{enumerate}
\end{definition}

These production rules allow to transform a sequence $\s$ of symbols into a 
sequence $\e$ of bits by successive applications of production rules. 
These rules are assumed to be reversible: inverting the direction
of the arrow allows to recover a given sequence $\s$ from 
the corresponding bitstream $e$. 
Note that a given production rule absorbs a symbol ($a_i$) 
and some bits ($\l_{i,j}$) 
from the temporary term to be encoded, 
and generates a given sequence of bits ($\b_{i,j}$).
Huffman codes are covered by this definition. 
More generally, a VLRS is a Fixed-to-Variable (F-to-V) Length code if
$\forall i\ |\Ri|=1 \text{ and } \l_i=\{\voidsequence\}$. 

\begin{example}
\label{ex:C1}
Code $\C_1=\{0,10,11\}$ can be seen as the following VLRS:
\begin{equation}
\nonumber
\begin{array}{lccc}
\R_{1,1}: & a_1 & \rightarrow & 0  \\
\R_{2,1}: & a_2 & \rightarrow & 10 \\
\R_{3,1}: & a_3 & \rightarrow & 11 \\
\end{array}
\end{equation}
\end{example}

Note that Definition~1 does not warranty that such a system leads 
to a valid prefix code. 
For example, a rule $\R_{i,j}$ where $\b_{i,j}$ is a prefix of $\l_{i,j}$ is not valid. 
In this paper, we focus on VLRS leading to valid codes.
Note that Suffix-constrained Codes introduced in \cite{JG04c} 
form a subset of VLRS and are characterized as follows.

\begin{definition}
\label{def:suffixconstrained}
A suffix-constrained code is a VLRS 
such that $\forall i, j\ \l_{i,j}$ is a suffix of $\b_{i,j}$. 
\end{definition}
\\

\begin{example} 
\label{ex:C2}
The following VLRS $\C_2$ is a suffix-constrained code:
\begin{equation}
\nonumber
\begin{array}{lccc}
\R_{1,1}: & a_1 0 & \rightarrow & 10  \\
\R_{1,2}: & a_1 1 & \rightarrow & 01  \\
\R_{2,1}: & a_2 & \rightarrow & 00 \\
\R_{3,1}: & a_3 & \rightarrow & 11 \\
\end{array}
\end{equation}
\end{example}

Note that Code $\C_2$ can not be encoded in the forward direction. 
We will come back on this point in Section~\ref{sec:encodingdecoding}.
The two following codes will also be considered in the sequel. 
Note that these codes are not suffix-constrained codes.

\begin{example} $\C_3$ is defined as
\label{ex:C3}
\begin{equation}
\nonumber
\begin{array}{lccc}
\R_{1,1}: & a_1 & \rightarrow & 00 \\
\R_{2,1}: & a_2 0 & \rightarrow & 01  \\
\R_{2,2}: & a_2 1 & \rightarrow & 10  \\
\R_{3,1}: & a_3 & \rightarrow & 11 \\
\end{array}
\end{equation}
\end{example}

\begin{example} $\C_4$ is defined as
\label{ex:C4}
\begin{equation}
\nonumber
\begin{array}{lccc}
\R_{1,1}: & a_1 1 & \rightarrow & 0  \\
\R_{1,2}: & a_1 0 & \rightarrow & 10  \\
\R_{2,1}: & a_2 & \rightarrow & 110 \\
\R_{3,1}: & a_3 & \rightarrow & 111 \\
\end{array}
\end{equation}
\end{example}

\begin{figure}[t]
\begin{center}
    \includegraphics[width=8cm]{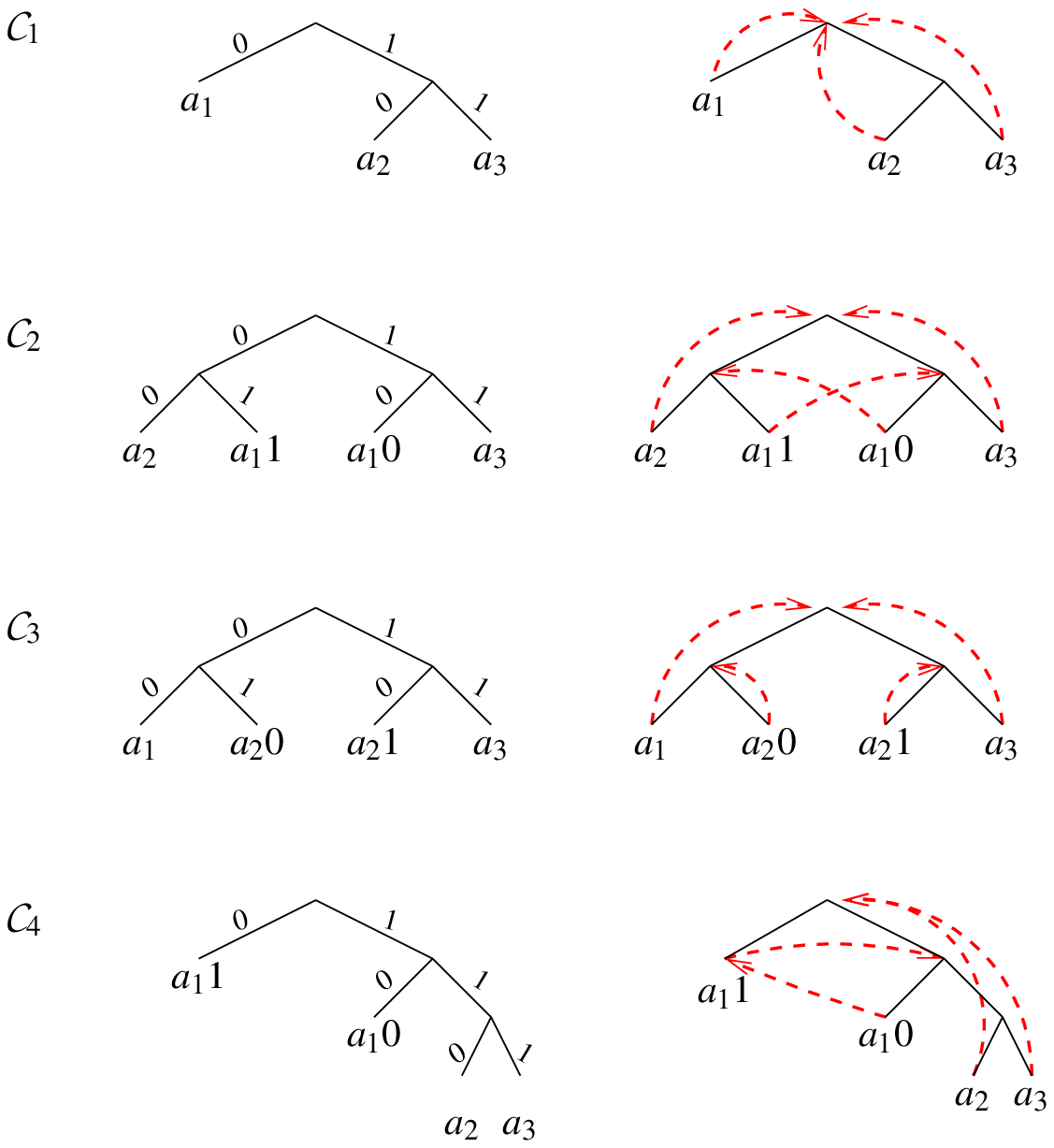}
\end{center}
\caption{Examples of VLRS: a VLC $\C_1$ and a suffix-constrained code $\C_2$. 
On the right, the transitions triggered by the production rules are depicted by arrows.}
\label{fig:exvlrs}
\end{figure}

VLRS can also be represented using trees, as depicted in Fig.~\ref{fig:exvlrs}. 
The tree structure corresponds to the one of the prefix code defined 
by $\bigcup_{i=1}^{|\A|} \bigcup_{j=1}^{|\Ri|} \{ \b_{i,j} \}$. 
Leaves correspond to both the symbol $a_i$ and the 
sequence of bits $\l_{i,j}$.

\section{Encoding and Decoding with automata}
\label{sec:encodingdecoding}

On the encoder side, the purpose of production rules is 
to transform the sequence ${\mathbf s}$ into the sequence ${\mathbf e}$ of bits.
Any segment of the current sequence (composed of 
symbols \emph{and} bits, initialized by ${\mathbf s}$)
can be rewritten if there exists a rule having this segment as an input
(this input is composed of one symbol \emph{and} a variable number of bits).
When the production rules stop, the sequence contains only bit entities.
The set of rules defining a VLRS does not generally allow 
to encode the sequence $\S$ in the forward direction. 
Therefore, the encoding must be processed backward. 
To initiate the encoding process a specific rule must be used to encode 
the last symbol of the sequence. 
Indeed the last symbol may not be sufficient to trigger a production
rule by itself. 
In most cases, they can be arbitrarily defined assuming 
that missing bit(s) equal ${\it 0}$, at the condition that the 
termination bit(s) do(es) not trigger a production rule. 
Hence, the choice $0$ is valid for the codes $\C_1$, $\C_2$ and $\C_3$ but should 
not be used for code $\C_4$, since $0$ triggers the rule $\R_{1,1}$. 

\begin{example}
\label{ex:encC1}
Let ${\mathbf s_1}=a_1 a_2 a_2 a_3 a_2 a_1 a_1 a_1$ be a sequence of symbols 
taking their values in the alphabet $\A_1=\{a_1,a_2,a_3\}$. 
This sequence is encoded with Code $\C_2$.
Since the last symbol is $a_1$, no rule applies directly. 
Therefore, the termination bit $0$ is concatenated 
to this sequence in order to initiate the encoding. 
The encoding then proceeds as follows:

\begin{center}
\begin{tabular}{ll}
$\R_{1,1}: $  & \hfill ${\mathbf s_1}{\it 0} = a_1 a_2 a_2 a_3 a_2 a_1 a_1 \underline{a_1 {\it 0}} $ \\
$\R_{1,2}: $  & \hfill $a_1 a_2 a_2 a_3 a_2 a_1 \underline{a_1 1} {\it 0}$ \\
$\R_{1,1}: $  & \hfill $a_1 a_2 a_2 a_3 a_2 \underline{a_1 0} 1 {\it 0} $ \\
$\R_{2,1}: $  & \hfill $a_1 a_2 a_2 a_3 \underline{a_2} 1 0 1 {\it 0}  $ \\
$\R_{3,1}: $  & \hfill $a_1 a_2 a_2 \underline{a_3} 0 0 1 0 1 {\it 0}  $ \\
$\R_{2,1}: $  & \hfill $a_1 a_2 \underline{a_2} 1 1 0 0 1 0 1 {\it 0}  $ \\
$\R_{2,1}: $  & \hfill $a_1 \underline{a_2} 0 0 1 1 0 0 1 0 1 {\it 0} $ \\
$\R_{1,1}: $  & \hfill $\underline{a_1 0} 0 0 0 1 1 0 0 1 0 1 {\it 0}$ \\
~             & \hfill ${\mathbf e_1} = 1  0 0 0 0 1 1 0 0 1 0 1 {\it 0} $ \\
\end{tabular}
\end{center}
\end{example}

In \cite{JG04c}, it was shown that transmitting the termination bit 
is not required for suffix-constrained codes, as shown in Example~\ref{ex:encC1}. 
This is due to the fact that a bit generated by a production rule 
of a suffix-constrained code will not be modified 
by a subsequent production rule.
Since these termination bits may be required in the general case, 
it will be assumed that they are known at the decoder. 
In the following example, the termination bit must be $1$. 
Note that the sequence is encoded with less than 1 bit per symbol. 

\begin{example}
\label{ex:encC4}
Let us now consider the sequence ${\mathbf s_1'}=a_1 a_1 a_1 a_1 a_1$. 
This sequence is encoded with code $\C_4$ as 

\begin{center}
\begin{tabular}{ll}
$\R_{1,1}: $  & \hfill ${\mathbf s_1'}{\it 1} = a_1 a_1 a_1 a_1  \underline{a_1 {\it 1}} $    \\
$\R_{1,2}: $  & \hfill $ a_1 a_1 a_1 \underline{a_1 0} $                                      \\
$\R_{1,1}: $  & \hfill $ a_1 a_1 \underline{a_1 1} 0 $                                        \\
$\R_{1,2}: $  & \hfill $ a_1 \underline{a_1 0} 0 $                                            \\
$\R_{1,1}: $  & \hfill $ \underline{a_1 1} 0 0 $                                              \\
~             & \hfill $ {\mathbf e_1'} = 0 0 0$                                        \\
\end{tabular}
\end{center}
\end{example}

On the decoder side, the decoding is processed forward using reverse rules. 
The encoding and decoding algorithms are implemented using automata. 
These automata are used to catch the memory of the encoding and decoding processes.
This memory corresponds to a segment of bits that may be useful for 
the next production rule. 
Hence, they are obtained directly from the set of production rules. 
The transitions on the automaton representing the encoding process 
are triggered by symbols. The internal states of the automaton are given 
by the variable length segments of bits $\{ \l_{i,j} \}$. 
This automaton may be reduced if a variable length bit segment $\l_{i,j}$ 
is a prefix of another segment $\l_{i',j'}$ 
(in that case, according to Definition~\ref{def:vlrs}, we have $i \neq i'$). 
If $\forall i,j,\ \l_{i,j}=\voidsequence$, 
there is only one internal state $\{ \voidsequence \}$ 
for the encoding automaton corresponding to code $\C_1$. 
The sets of states of encoding automata of codes $\C_2$, $\C_3$
and $\C_4$ are identical and are equal to $\{ 0, 1 \}$.

The states of the decoding automata correspond to bit segments
that have already been decoded, but which are not sufficient to identify 
a symbol. For VLCs such as Huffman codes, these internal states
correspond to the internal nodes of the decoding codetree. 

\begin{example}
The set of internal states of codes $\C_1$, $\C_2$, $\C_3$ and $\C_4$ are respectively 
$\{ \voidsequence,1 \}$,
$\{ \voidsequence,0,1 \}$,
$\{ \voidsequence,0,1 \}$ and
$\{ \voidsequence,1,11 \}$.
\end{example}

\begin{figure}[t]
\begin{center}
    \includegraphics[width=8cm]{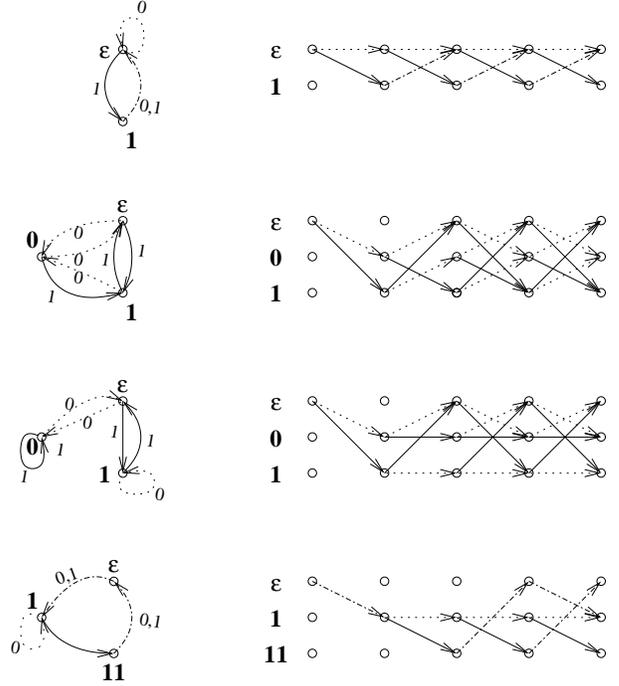}
\end{center}
\caption{Decoding automata corresponding to the codes $\C_1$, $\C_2$, $\C_3$ and $\C_4$
and corresponding decoding trellises. 
Transitions corresponding to 0{\it s} and 1{\it s} 
are respectively plotted with dotted and solid lines.}
\label{fig:decauto}
\end{figure}

The graphical representations of the decoding automata may be deduced 
from the tree representations given in Fig.~\ref{fig:exvlrs}. 
These automata are depicted in Fig.~\ref{fig:decauto}.
The decoding trellises corresponding to these automata are depicted on the right. 
For sake of clarity, the symbols generated by the bit transitions are not shown.
However, note that the set of generated symbol(s) must also be associated 
to each bit transition. 
For the codes $\C_1$, $\C_2$ and $\C_3$, 
at most 1 symbol is associated to each bit transition. 
It is not the case for Code $\C_4$, where the transition 
starting from decoding state ${\bf 1}$ 
triggered by the bit $0$ generates the symbol $a_1$ twice. 
As shown in Example~\ref{ex:encC4} and demonstrated 
in Section~\ref{sec:compressionefficiency}, this transition 
allows to encode long sequences of $a_1$ with less than $1$ bit, 
at the cost of a higher encoding cost for the symbols $a_2$ and $a_3$. 


\section{Compression efficiency}
\label{sec:compressionefficiency}

In this section, we analyse the compression efficiency of VLRS{\it s}. 
Let us assume that $\S$ is a memoryless source characterized by 
its stationary probability distribution function (pdf) 
on $\A$: $\boldsymbol{\mu}=\{ \P(a_1), \hdots \P(a_i), \hdots \}$. 
Let 
\begin{equation}
\delta(\R_{i,j}) = L(\b_{i,j})-L(\l_{i,j})
\end{equation}
denote the number of bits generated by a given production rule $\R_{i,j}$. 
Note that for the particular case where 
$\forall i,\ \forall j, j'\ \delta(\R_{i,j})=\delta(\R_{i,j'})$, 
the mdl is equal to $\sum_{a_i \in \A} \P(a_i) \delta_{i,1}$. 

\begin{example}
\label{ex:mdl1}
Let us assume that $\S$ is a memoryless source of pdf
$\boldsymbol{\mu}_1=\{0.7, 0.2, 0.1\}$. The entropy 
of this source is $1.157$. The mdl of Code $\C_1$ is equal to $1.3$. 
For the code $\C_2$, we have $\delta(\R_{1,1})=\delta(\R_{1,2})=1$ 
and $\delta(\R_{2,1})=\delta(\R_{3,1})=2$. The mdl of this code 
is also equal to $1.3$. 
\end{example}

Let $R_t: S_t \overline{L}_t \rightarrow \overline{B}_t $ 
denote the rule to be used in order to encode a given symbol $S_t$.
Since the encoder proceeds backward and since the source $\S$
is memoryless, the process $(Z_{t'})=(R_{L(\S)}, \hdots R_{L(\S)}, \hdots R_{1})$ 
obtained from the process $(R_t)$ by reversing the symbol clock $t$, 
i.e. $(Z_{t'})_{t'=1,\hdots L(\S)}=(R_{L(\S) - t + 1})_{t=1,\hdots L(\S)}$, 
forms an invariant Markov chain. 
In other words we have 
$\P(Z_{t'} | Z_1, \hdots Z_{t'-1}) = \P(Z_{t'} | Z_{t'-1}) 
= \P( R_t | R_{t+1}, \hdots R_{L(\S)}) = \P( R_t | R_{t+1} ) = \P( R_{L(\S)-1} | R_{L(\S)})$. 
If $S_t=a_i$, the rule $\R_{i,j}$ is triggered if and only if the realization of 
$\overline{L}_t$ is a prefix of the bits $\overline{B}_{t+1}$ generated 
by the previous production rule. 
As a consequence, the probability $\P( R_t | R_{t+1} )$ 
can be deduced from the source pdf as 
\begin{align}
\P( R_t=\R_{i,j} | & R_{t+1} = \R_{i',j'} ) \\
& = \P( R_t=\R_{i,j} | \overline{B}_{t+1}=\b_{i',j'} ) \nonumber \\
& = \P( S_t = a_i, \overline{L}_t=\l_{i,j} | \overline{B}_{t+1}=\b_{i',j'} ) \nonumber \\
& = \left\{
\begin{array}{ll}
\P(a_i) & \text{ if $\l_{i,j}$ 
is prefix of $\b_{i',j'},$ } \nonumber \\
0 & \text{ otherwise.}
\end{array}
\right.
\end{align}

Assuming that $(Z_{t'})$ is irreducible and aperiodic,
the marginal probability distribution
$\P(Z_{t'}=\R_{i,j})$ is obtained from the transition matrix 
$\P(Z_{t'} | Z_{t'-1})$ as the normalized eigenvector 
associated to the eigenvalue~1. 
As $t'$ grows to infinity (which requires that $t \rightarrow \infty$), 
the expectation of $\delta(Z_{t'})$ is the expectation of 
the number of bits generated by a production rule. With the Cesaro theorem, 
it also provides the asymptotic value of the mdl as the sequence length increases.

\begin{example}
\label{ex:mdlC4}
For the code $\C_4$, the transition matrix corresponding to 
the source pdf of Example~\ref{ex:mdl1} is  
$$
\left[
\begin{array}{cccc}
0 & 0.7 & 0.7 & 0.7 \\
0.7 & 0 & 0 & 0 \\
0.2 & 0.2 & 0.2 & 0.2 \\
0.1 & 0.1 & 0.1 & 0.1 \\
\end{array}
\right], 
$$
which leads to $\P(R_t=r_{i,j}) = \{ 0.412,0.288,0.2,0.1 \}$. 
Finally, the mdl of this code is 
$mdl(\C_4)=0.412 \times 0+0.288 \times 1+0.2 \times 3+0.1 \times 3=1.188$. 
\end{example}

The mdl obtained in Example~\ref{ex:mdlC4} is much closer to the 
entropy than the mdl obtained with Huffman codes. 
The expected number of bits required to code the symbol $a_1$ 
is less than 0.5 bit.
One can also process the exact mdl of a VLRS for sequences of finite length.
Indeed, the expectation of the number of termination bit(s) 
as well as the pdf $\P(R_t=r_{i,j} | t = L(\S))$ of the last rule can be obtained 
from the termination bit choice and from the source pdf. 
The exact probability $\P(R_t=r_{i,j}|t=\tau)$ of having a given rule 
for a given symbol clock $\tau$ can then be computed 
and subsequently one can deduce the expectation of the number of bits generated
to encode the symbol $S_{\tau}$. 

\section{Lexicographic Code Design}
\label{sec:lexi}

This section describes a VLRS construction method which allows to preserve 
the lexicographic order of the source alphabet in the bit domain. 
As a starting point, we assume that the Huffman 
code corresponding to the source pdf $\boldsymbol{\mu}$ is already known. 
The length of the Huffman codeword associated to the symbol $a_i$ is denoted $k_i$. 
Let $k^+=\max_i k_i$ denote the length of the longest codeword. 
First, let us underline that the union $\bigcup_{i,j} \{ \b_{i,j} \}$ 
of all the bit sequences $\b_{i,j}$ will form a Fixed Length Code (FLC) 
${\mathcal F}$ of length $k^+$. ${\mathcal F}$ contains $2^{k^+}$ codewords. 
These codewords will be assigned to productions rules in the lexicographic order. 
Starting with the smaller symbol $a_1$, 
$2^{k^+ - k_i}$ rules are defined for symbol $a_i$.
The left part of these rules are defined so that 
the set $\{\l_{i,j} \}_{j \in [1..|\Ri|]}$ forms a FLC of length $k^+-k_i$. 
If $k_i=k^+$, this FLC only contains the element $\voidsequence$. 
The $2^{k^+ - k_i}$ smallest remaining codewords of ${\mathcal F}$, i.e. those which 
have not been assigned to previous symbols of ${\mathcal F}$, 
are then assigned to these productions rules so that 
$\forall j, \l_{i,j} \leq \l_{i,j'} \Rightarrow \b_{i,j} \leq \b_{i,j'}$. 
By construction, the proposed algorithm leads to a VLRS with the 
lexicographic property and with the same compression efficiency as the code
from which it is constructed. 
In some cases, 
the set of production rules generated in previous steps may be simplified. 

\begin{example}
\label{ex:lexi1}
Let us now assume that the source ${\mathbf S}$ is memoryless 
of pdf $\boldsymbol{\mu}_2 = \{ 0.2,0.7,0.1 \}$. 
Since $a_2$ has the highest probability, the Huffman code ${\mathcal H}_2=\{10,0,11 \}$ 
corresponding to this pdf is not lexicographic.
The Hu-Tucker code associated to this source is the code $\C_1$ 
proposed in Example~\ref{ex:C1} and its mdl is equal to $1.8$.

The VLRS is constructed according to the proposed construction procedure. 
For ${\mathcal H}_2$, we have $k_2=1$ and $k_1=k_3=k^+=2$. 
Hence ${\mathcal F}=\{00,01,10,11\}$. Since $k_1=2$, only $1$ production 
rule $r_{1,1}$ is assigned to the symbol $a_1$ and $b_{1,1}=\voidsequence$, 
which implies $r_{1,1}: a_1 \rightarrow 00$. 
The symbol $a_1$ is then assigned two production rules $r_{2,1}$ and $r_{2,2}$ 
so that $r_{2,1}: a_2 0 \rightarrow 01$ and $r_{2,2}: a_2 1 \rightarrow 10$. 
The construction algorithm finishes with the assignment 
of rule $r_{3,1}:a_3 \rightarrow 11$ to symbol $a_3$. 
Finally, we obtain the code $\C_3$ proposed in Example~\ref{ex:C3}, 
for which the mdl is equal to $1.3$ together with the lexicographic property. 
\end{example}

Although the proposed construction allows to obtain lexicographic codes with 
the same compression efficiency as Huffman codes, 
it does not construct, in general, the best lexicographic VLRS 
from a compression efficiency point of view. 
One may find some lexicographic VLRS with lower mdl. 

\section{Mirror Code Design}
\label{sec:mirror}

\begin{figure}[t]
\begin{center}
    \includegraphics[width=8cm]{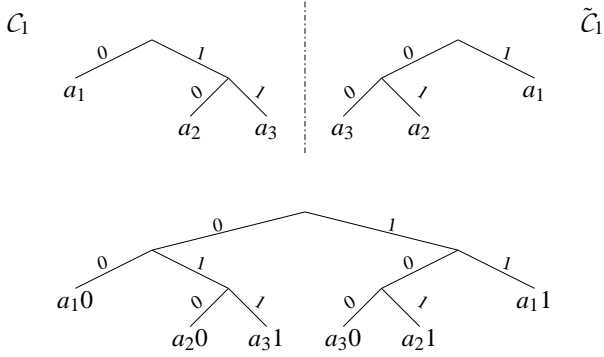}
\end{center}
\caption{Primitive code $\C_1$, its opposite $\tilde{\C}_1$ 
and the resulting mirror VLRS.}
\label{fig:mirror}
\end{figure}

The code design described in this section allows to obtain codes with
bit marginal probabilities that are asymptotically equal to 0.5 
as the sequence length increases. 
Let us again assume, as a starting point, that we know a VLC code 
${\mathcal H}=\{ \b_{1,1}, \hdots \b_{|\A|,1}\}$. 
Let us now consider the code ${\mathcal \tilde{H}}=\{\tilde{\b}_{1,1}, \hdots \tilde{\b}_{|\A|,1} \}$ 
defined so that each bit transition of the codetree characterizing ${\mathcal \tilde{H}}$ 
is the opposite value from the corresponding bit transition in ${\mathcal H}$,
as depicted in Fig.~\ref{fig:mirror}. 

The VLRS is obtained by putting together these two codes. 
The codes ${\mathcal H}$ and $\tilde{{\mathcal H}}$ are respectively 
used to define the two sets of $|\A|$ production rules forming the 
new VLRS as 
\begin{equation}
{\mathcal M}=
\left\{
\begin{array}{l}
\{ a_i b_{i,1}^{L(\b_{i,1})} \rightarrow 0 \, \b_{i,1} \}_{i \in [1..|\A|]} \\
\{ a_i \tilde{b}_{i,1}^{L(\tilde{\b}_{i,1})} \rightarrow 1 \, \tilde{\b}_{i,1} \}_{i \in [1..|\A|]}.
\end{array}
\right.
\end{equation}
Note that the production rules associated 
to codes ${\mathcal H}$ and ${\mathcal \tilde{H}}$ 
respectively define the subtrees corresponding to bit transitions $0$ and $1$. 
Note also that the resulting code, by construction, is a suffix-constrained code.

\begin{example}
\label{ex:mirror} 
The construction associated to the code $\C_1$ leads to the following VLRS:
\begin{equation}
\nonumber
\underbrace{
\left\{
\begin{array}{lccc}
\R_{1,1}: & a_1 0 & \rightarrow & 00  \\
\R_{2,1}: & a_2 0 & \rightarrow & 010 \\
\R_{3,1}: & a_3 1 & \rightarrow & 011 
\end{array}
\right.
}_{\text{obtained from } {\mathcal H}=\C_1}
\hfill
\underbrace{
\left\{
\begin{array}{lccc}
\R_{1,2}: & a_1 1 & \rightarrow & 11  \\
\R_{2,2}: & a_2 1 & \rightarrow & 101 \\
\R_{3,2}: & a_3 0 & \rightarrow & 100. 
\end{array}
\right.
}_{\text{obtained from } \tilde{\mathcal H}=\tilde{\C}_1}
\end{equation}
\end{example}

{\vspace{-0.6cm} \noindent \it Proof of 
$\forall n,\ \lim_{L(\S) \rightarrow \infty} \P(E_n=0)=0.5$: } 
Let us consider a VLRS ${\mathcal M}$ constructed according to the previous 
guidelines. The notation $b_{i,j}$ refers to this VLRS (not to the VLC from 
which it is constructed). 
Let $f_t=\P(B_t^1 = 0)$ denote the marginal bit probability associated to 
the first bit generated by a given production rule. 
Since the VLRS is constructed from a VLC, we have 
$\forall i,j,\ \delta(\R_{i,j}) \geq 1$,
which means that every rule produces at least one bit. 
The value $f_t$ can be written as 
\begin{align}
  f_t & = \sum_{i \in [1..|\A|], j \in [1..2]} \P(R_t=\R_{i,j}, B_t^1 = 0 )  \\
  ~ & = \sum_{i \in [1..|\A|], j=1} \P(S_t=a_i, B_{t+1}^1= l_{i,1}^{L(l_{i,1})}) \\
  ~ & = \sum_{i \in [1..|\A|], j=1} \P(S_t=a_i, L_t^{L(L_t)} = 0 ) \, f_{t+1} \nonumber \\
  ~ & + \sum_{i \in [1..|\A|], j=1} \P(S_t=a_i, L_t^{L(L_t)} = 1 ) \,  (1-f_{t+1}).
\label{equ:ft1}
\end{align}
Let $\alpha=\sum_{a_i \in \A} \P( a_i,l_{i,1}^{L(\l_{i,1})} = 0 )$. 
This entity corresponds to the sum of the probabilities of the symbols to which 
a codeword ending with 0 has been assigned. Note that $0 < \alpha < 1$. 
Inserting this entity in Eqn.~\ref{equ:ft1}, we obtain 
\begin{align}
f_t = \alpha \, f_{t+1} + (1-\alpha) (1-f_{t+1}). 
\end{align} 
We can now study the asymptotic behavior of this sequence 
as $t'=L(\S)-t+1$ tends to $+\infty$ 
(note that $f_{L(\S)}$ is a constant). 
The absolute value of the derivative of the 
function $g(x)=\alpha \, x + (1-\alpha) (1-x)$ 
is strictly lower than 1 when $0 < \alpha < 1$. 
Consequently, the fixed-point theorem applies and 
the sequence $f_{L(\S)}, f_{L(\S)-1}, \hdots f_{t'}$ 
converges to the solution of 
$x = g(x)$, which is 0.5. 
Subsequently, $\forall i$, 
opposite codewords $\b_{i,1}$ and $\b_{i,2}$ are equiprobable, 
which concludes the proof.
\hfill $\Box$


\section{Conclusion and Perspectives}

VLRS{\it s} have a low encoding and decoding complexity, allowing for instantaneous 
decoding and may have a lower mdl than Huffman codes. 
The degree of freedom that they offer allows to design codes with 
interesting properties, as shown in sections \ref{sec:lexi} and \ref{sec:mirror}. 
Hopefully, the design of \ref{sec:mirror} may lead 
to soft decoding results outperforming the ones obtained with source codes 
with a marginal bit probability not equal to 0.5. 
 

\vspace{-0.3cm}
\bibliographystyle{IEEEtran}
\bibliography{abrv,JG,biblio,softVLC}

\end{document}